\documentstyle[aps,prl,floats]{revtex}
\begin{document}
\draft
\input epsf.sty

\twocolumn[\hsize\textwidth\columnwidth\hsize\csname 
@twocolumnfalse\endcsname
\title{Optically Pumped NMR Studies of Electron Spin 
Polarization and Dynamics:\newline
New Constraints on the Composite Fermion Description 
of {\boldmath $\nu$} = 1/2}
\author{A.\,E. Dementyev$^{1}$, N.\,N. Kuzma$^{1}$, 
P. Khandelwal$^{1}$, S.\,E. Barrett$^{1}$, 
L.\,N. Pfeiffer$^{2}$, and K.\,W. West$^{2}$}
\address{$^{1}$Department of Physics, 
  Yale University, New Haven, Connecticut 06511 
\\ $^{2}$Bell Laboratories, Lucent Technologies, 
  Murray Hill, New Jersey 07974}
\date{\today}
\maketitle
\begin{abstract}
 The Knight shift and the nuclear spin-lattice
relaxation rate 1/T$_1$
of $^{71}$Ga nuclei are measured in 
GaAs  quantum wells at Landau level filling factor
 $\nu$ = 1/2 using optically pumped NMR. 
 The temperature dependences of these data  are compared with 
predictions of a 
weakly-interacting  composite fermion model.  Our measurements
of the electron spin polarization and spin dynamics 
 near the transition between fully 
and partially spin-polarized ground states
provide new constraints
on the theoretical description of $\nu$ = 1/2.

\end{abstract}

\vskip 2pc ] 

\narrowtext

Significant progress in describing  a strongly 
interacting 2-dimensional electron system (2DES) 
in a large  magnetic field in terms of  
composite fermions (CF) in a reduced field has 
stimulated a large body of 
theoretical and experimental work 
\cite{cfrevs,rectheory}. One of the most suprising 
implementations of this idea was put forth in the 
seminal work of Halperin, Lee, and Read \cite{HLR}, 
who argued that the ground state of the 2DES at 
Landau level filling factor $\nu$=$\frac{1}{2}$ is 
well-described by CF in zero net magnetic field, 
which therefore exhibit a well-defined Fermi 
surface. Experiments carried out  near $\nu$=$
\frac{1}{2}$ have provided convincing evidence of 
the existence of the CF Fermi surface
\cite{willett93,willett95,kang,goldman94,smet}.

Despite the overall agreement between theory and 
experiment to date, several fundamental issues 
about CF at $\nu$=$\frac{1}{2}$ have yet to be 
resolved experimentally. For example, do CF form 
a Fermi gas, a ``normal'' Fermi liquid, or some 
kind of ``unusual'' Fermi liquid?  Also, 
does the picture change 
when the ground state is only partially 
\mbox{spin-polarized\cite{du,kukushkin,polarmass}?}
  Experiments which directly 
probe the electron spin degree of freedom  
{\em right at} $\nu$=$\frac {1}{2}$,
especially near the transition between partially 
and fully spin-polarized ground states,
will help to answer these central questions. 

In this Letter, we report optically pumped nuclear 
magnetic resonance (OPNMR) \cite{opnmrchar} 
measurements of the Knight Shift $K_S$ and 
the spin-lattice relaxation rate 1/T$_1$ 
of $^{71}$Ga  nuclei in two different electron-doped 
multiple quantum well (MQW) samples. The $K_S$ 
data reveal the  spin polarization ${\cal P}(T)
$\,$\equiv$\,$\frac{\langle S_{z}(T)\rangle}
{\text{max}\langle S_{z}\rangle}$, while
the 1/T$_1$ data probe the spin dynamics of the 
2DES. Taken together, these 
thermodynamic measurements provide unique insights 
into the physics of CF at $\nu$=$\frac{1}{2}$.

The two samples used in this work were previously 
studied\cite{pankaj,nick}
near $\nu$=$\frac{1}{3}$.  Sample  40W contains 
forty $300\,$\AA\space wide GaAs wells separated
by $3600\,$\AA\space wide Al$_{0.1}$Ga$_{0.9}$As 
barriers.   Sample 10W contains ten
$260\,$\AA\space wide wells separated by  
$3120\,$\AA\space wide  barriers.  Silicon 
delta-doping spikes located in the center of 
each barrier  provide the electrons that 
are confined in each GaAs well at low temperatures, 
producing 2DES with very high mobility
($\mu > 1.4\times10^{6}$~cm$^{2}$/Vs).  This MQW 
structure  results in a 2D electron density that 
is unusually insensitive to light, and extremely 
uniform from well to well\cite{insensitive}.  
The low temperature ($0.29\,<$$\,T$$\,<\,$$
3.5\,$K) OPNMR measurements described below were 
performed using  a sorption-pumped~$^{3}$He 
cryostat.  The samples, about \mbox{
$4\,$$\times\,$6$\,$mm$^{2}$} in size, were in 
direct contact with $^{3}$He, mounted on the 
platform of a rotator assembly in the NMR probe.  
Data were acquired following the previously 
described\cite{pankaj,nick,opnmr,science} OPNMR 
timing sequence:
\mbox{SAT--$\tau_{L}$--$\tau_{D}$--DET}, modified 
for use below 1 Kelvin (e.g., $
\tau_{D}\,$$>\,$10\,s, laser power 
\mbox{$\sim10\,$mW/cm$^{2}$}, low rf voltage 
levels). A calibrated RuO$_{2}$ sensor, in 
good thermal contact with the sample, was used to
monitor the temperature.
The peak in $K_S(\nu)$ at $\nu$=1 was used to 
determine the electron density $n$ for each 
sample\cite{pankaj}. 
Using the rotator assembly, we could set the angle 
$\theta$ between the sample's growth axis and the 
applied field $B_{\text{tot}}$ so that
the filling factor $\nu$=$nhc/eB_{\bot}$  (with 
$B_{\bot}$$\equiv$$B_{\text{tot}}\!\cos\theta$) equalled $\frac{1}{2}$
  for these three cases:\vspace{2mm}\newline
\begin{tabular}{c|ccrcc}
 \space Case \space 
          & \hspace{1mm}Sample\hspace{1mm} & 
$B_{\text{tot}}$ [T] & \multicolumn{1}{c}{$\theta$}
 & $n$ [$\text{cm}^{-2}$] & $w$ [\AA] \\ \hline
I   & 40W & 7.03 & \hspace{1mm} 38.3$^{\circ}$ 
\hspace{1mm} & \hspace{1mm}
6.69$\times\,$$10^{10}$ \hspace{1mm} & 300 \\ 
II  & 40W & 5.52 & 0.0$^{\circ}$ \hspace{1mm} &
 6.69$\times\,$$10^{10}$ & 300 \\ 
III  & 10W & 7.03 & 24.5$^{\circ}$ \hspace{1mm} & 
7.75$\times\,$$10^{10}$ & 260 \\ \hline
\end{tabular}

\begin{figure}
\centerline{\epsfxsize=3.0in\epsfbox{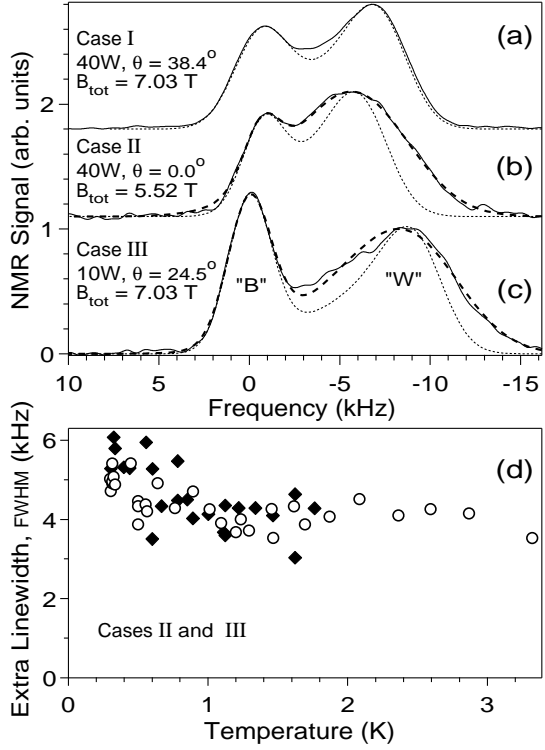}}
\caption{$^{71}$Ga OPNMR spectra (a-c, solid lines) 
at $\nu$=$\frac{1}{2}$, $T$\,$\approx$\,0.5\,K. 
The dotted line fits (a-c) use a 3.5 Khz FWHM
Gaussian broadening (due to nuclear spin-spin
coupling) of the intrinsic well and barrier
lines. The dashed line fits (b,c) require
an extra Gaussian broadening of  ``W''
with the FWHM shown in (d) for 
samples 40W (open circles) and  10W (filled 
diamonds).
}
\label{fig1}
\end{figure}

\vspace{2mm}Figure \ref{fig1} shows  OPNMR spectra   
at $\nu$=$\frac{1}{2}$ and $T$\,$\approx$\,$0.5$\,K, 
for Cases I--III (a-c, solid lines).
Nuclei within
the quantum wells are coupled to the spins of the 
2DES via the isotropic Fermi contact 
interaction\cite{slichter}, which shifts the 
corresponding well resonance (labeled ``W" on 
Fig.\,\ref{fig1}(c))  relative to the signal from 
the barriers (``B")\cite{opnmr,science}.
We define the Knight shift $K_S$ as the 
peak-to-peak splitting between ``W" 
and ``B". For Case I, all spectra (e.g., 
Fig.\,\ref{fig1}(a)) are well-described 
by the same two-parameter 
fit (dotted lines)\cite{pankaj,nick} that was used 
for all spectra at $\nu$=$\frac{1}{3}$.  
This fit is generated under the assumption that 
all spins are delocalized, so that 
$\langle S_{z}(\nu,T)\rangle$, averaged over the 
NMR time scale ($\sim\,$40$\,\mu$sec), appears 
spatially homogeneous along the plane of the wells, 
and thus the resulting lineshape is 
``motionally-narrowed'' \cite{slichter}.

In contrast, for Cases II and III, the well 
resonance (Fig.\,\ref{fig1}(b,c))
is much broader than the same fit (dotted lines).  
An additional gaussian broadening of just the  
well resonance leads to a better fit (dashed 
lines). The full width at half maximum (FWHM) of 
the additional broadening extracted from these 
fits is plotted  in 
Fig.\,\ref{fig1}(d) for Cases II and III. 
Earlier measurements at
$\nu$\,$<$\,$\frac{1}{3}$ and
$T$$\approx$\,0.5\,K were also poorly
described by the ``motionally-narrowed'' lineshape, 
but in that case the extra well width was sharply 
temperature-dependent \cite{nick}, whereas here
it is essentially independent of temperature. 
The extra 
broadening of the well lineshape for Cases II and 
III (Fig.\,\ref{fig1}(b,c,d)) seems to be 
homogeneous \cite{density}, if so, the corresponding 
 transverse relaxation time T$_2$ \cite{slichter} is
quite short. The origin 
of this effect is not understood as it is very
hard to explain simultaneously 
the temperature-independence of the
 extra broadening  and the lack of a 
similar effect in Case I.

\begin{figure}
\centerline{\epsfxsize=3.0in\epsfbox{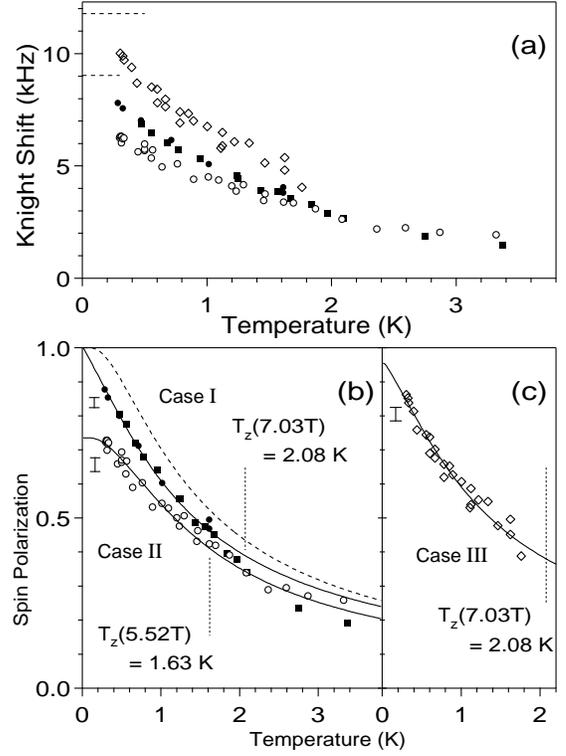}}
\caption{ 	
Temperature dependence at $\nu$=$\frac{1}{2}$ 
of (a) $K_S$ and (b, c) ${\cal P}$ for Case I 
(open diamonds), Case II (filled symbols),
and Case III (open circles).
Note  the error bars  in (b,c). 
The solid and dashed curves are 
described in the text.
}
\label{fig2}
\end{figure}

Figure \ref{fig2}(a) 
shows $K_S(T)$  at $\nu$=$\frac{1}{2}$ 
for Cases I--III.  The larger scatter in the $K_S(T)$ 
data for Cases II and III is a consequence of the 
large linewidth.
Using an empirical relation  (all in kHz)
$K_{S\text{int}}$\,= $K_S$$+$$1.1$$\times$$(1$$-$$
\exp(-K_S/2.0))$, we can convert $K_S$ into 
$K_{S\text{int}}$, which is the 
intrinsic hyperfine shift for
the nuclei in the center of each well. 
$K_{S\text{int}}\,$=$\,A_c\:{\cal P}\:n/w$ is a 
direct measure of the electron spin polarization 
${\cal P}$, where $w$ is the width of each well 
and $A_c=(4.5\pm 0.2)\times 10^{-13}\,$cm$^{3}$/s 
is the hyperfine constant\cite{pankaj}.
For Cases II and III, the same values of 
$K_{S\text{int}}(T)\,$ are also obtained
directly from the dashed line fits (e.g., 
Fig.\,\ref{fig1}(b,c)).

$K_{S\text{int}}(T)$ is converted to 
electron spin polarization using  
\mbox{${\cal P}(\nu${\small=}$\frac{1}
{2}$,$T)$}=$K_{S\text{int}}(T)/
K_{S\text{int}}^{{\cal P}=1}$, where 
the maximum shift for a fully polarized 2DES 
 is known for each sample:
$K_{S\text{int}}^{{\cal P}=1}$\,=
$K_{S\text{int}}(\nu${\small =}$\frac{1}{3},$$
T$$\rightarrow$0) \cite{pankaj}.

Figure \ref{fig2}(b,c) shows that 
${\cal P}(\nu${\small=}$\frac{1}{2},T)$ does not 
saturate down to our base temperature of 
0.29\,K, in contrast to earlier measurements 
at $\nu$=1 and $\frac{1}{3}$ \cite{pankaj,opnmr}. 
 Moreover, as the temperature is 
increased, ${\cal P}(\nu,T)$ falls off much 
faster at $\nu$=$\frac{1}{2}$
than at $\nu$=$1$ or $\frac{1}{3}$ (e.g., at 
$T_{Z}$=$|$g$^{\ast}\mu_e B_{\text{tot}}/k_B|$,
${\cal P}(\nu$=1,$T_{Z})$\,$\approx$\,$93\%$, 
while ${\cal P}(\nu$=$\frac{1}{2},$$T_{Z})$\,$
\approx$\,40$\%$. Here \mbox{g$^{\ast}$=--0.44,}
 $\mu_e$ is the Bohr
magneton and $k_B$ is the Boltzmann constant).
  Qualitatively, these results 
are consistent with a tiny (or vanishing)
energy gap for spin-flip excitations at 
$\nu$=$\frac{1}{2}$ for Cases I--III.
However, a quantitative understanding of the 
${\cal P}(\nu$=$\frac{1}{2},T)$ data remains a 
challenge for theory (e.g., we cannot explain
 the crossing of the Case I and Case II data sets 
 at $T$$\approx$$T_{Z}$ (Fig.\,\ref{fig2}(b))).

Even though saturation is not observed, the Knight
shift data for Case II are evidence for a
$\nu$=$\frac{1}{2}$ ground state in which the 
electrons are only {\em partially} spin-polarized
(i.e., ${\cal P}(\nu$=$\frac{1}{2},T\!\rightarrow0)
$\,$\approx$\,$70-85\%$).  This inference is consistent 
with data obtained from two other experiments in
 conditions 
similar to those of Case II.  From their time-resolved 
photoluminescence measurements, Kukushkin et al.\cite{kukushkin} 
estimate ${\cal P}(\nu$=$\frac{1}{2},T\!\rightarrow0)
$\,$\approx$\,$87\%$ at 5.52\,T. Surface acoustic wave 
measurements of Willett et al.\ obtained a Fermi wave 
vector which was $\sim90\%$ of the theoretical value
for  fully polarized CF, consistent with
a polarization of at least
${\cal P}(\nu$=$
\frac{1}{2},T\!\rightarrow0)\approx 62\%$.\cite{willett93}

The solid curves in Figure \ref{fig2}(b,c)  
are two parameter 
fits to the $T$$<$$T_Z$ data using expressions
for 
${\cal P}(\nu$=$\frac{1}{2},T)$ 
derived within a weakly-interacting 
composite fermion model (WICFM).  In this 
model, the dispersion relations for spin-up 
and spin-down  states
are:
\begin{equation}  
E_{\uparrow}(k)=\frac{\hbar^2k^2}{2m^{\ast}}, 
\,\,\,\, E_{\downarrow}(k)=\frac{\hbar^2k^2
}{2m^{\ast}} + E_Z^{\ast}(T)\,,
\label{eq1}
\end{equation}
where an  exchange 
interaction has been included in the model 
through the effective Zeeman energy:
\begin{equation}  
E_Z^{\ast}(T)=   |$g$^{\ast}\mu_eB_{
\text{tot}}|+     E_{\text{Exch}}=
k_B T_Z + k_B J {\cal P}(T)\,.
\label{eq2}
\end{equation}
When $J=0$, this is just the non-interacting 
composite fermion model\cite{kukushkin}. When $J>0$, there
is a Stoner enhancement of the spin 
susceptibility. In this model, the chemical potential 
$\mu$ and the polarization  ${\cal P}$ 
are:

\begin{equation}  
\mu(T)= 
k_BT\ln\Bigl(-\gamma
 + \sqrt{\gamma^2+
\exp(\rho)-1}\Bigr) + \frac{E_Z^{\ast}(T)}{2}
\label{eq3}
\end{equation}

\begin{equation}    
{\cal P}(T) = \frac{1}{\rho} \ln 
\Bigl(\frac{1+\exp[\frac{\mu(T)}{k_BT}]}
{1+\exp[\frac{\mu(T)}{k_BT}(1-\delta(T))]}
\Bigr) 
\label{eq4}
\end{equation}
\begin{displaymath}
{\text{where}}\,\,\,\,\,\gamma$=$
\cosh({\frac{E_Z^{\ast}(T)}{2k_BT}})$;\, 
$\rho$=$\frac{2\pi\hbar^2n}{m^{\ast} k_BT}$;\, and\,
$\delta(T)$=$\frac{E_Z^{\ast}(T)}{\mu(T)}.
\end{displaymath}

\begin{figure}
\centerline{\epsfxsize=3.0in\epsfbox{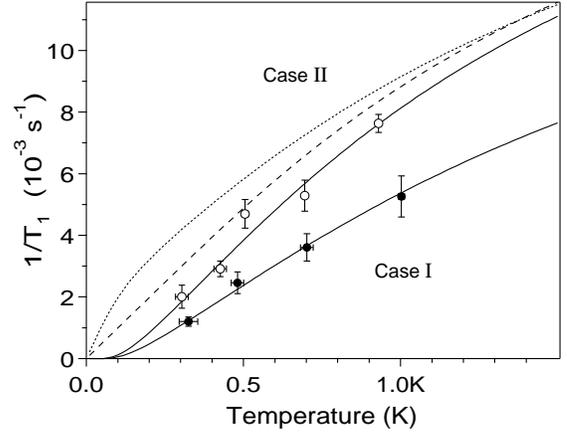}}
\caption{ Temperature dependence of the $^{71}
$Ga spin-lattice relaxation rate 1/T$_{1}$ at 
$\nu$=$\frac{1}{2}$ for Case I (filled symbols) 
and Case II (open symbols).
The solid and dashed curves are 
described in the text.}
\label{fig3}
\end{figure}

Equations 2--4 are solved 
self-consistently for ${\cal P}(T)$ at each $m^{\ast}$ and $J$.
 Within this WICFM, the behavior 
of ${\cal P}(T)$ as $T$$\rightarrow$0 is
quite sensitive to the parameter $\delta(0)$.
The ground state is only fully
polarized (${\cal P}(0)$=1) when $\delta(0)$$\geq$1. 
We find $\delta(0)$$<$1 for Cases II and III, 
$\delta(0)$\,=\,1 for Case I, and 
the dashed curve illustrates $\delta(0)$\,$>$\,1 
(Fig.\,\ref{fig2}(b,c)). 
Thus, within this model, the best-fit curves 
for Cases II and III yield partially-polarized
ground states, while Case I is fully-polarized.  
As described earlier, Cases II and III also have 
extra linewidth,
while Case I does not (Fig. \ref{fig1}).

Figure \ref{fig3} shows the temperature 
dependence of the $^{71}$Ga nuclear spin-lattice 
relaxation rate 1/T$_1$ at $\nu$=$\frac{1}{2}$
for Cases I and II.  At each temperature,
OPNMR spectra were acquired using a series of 
dark times 10\,s $\leq$ $\tau_{\text{D}}$ 
$\leq$ 2560\,s (i.e., the
longest $\tau_{\text{D}}$\,$\geq$\,$4T_1$).
The value of 1/T$_1$ was determined by fitting 
the signal intensity at the ``W'' peak frequency 
 to the form $S(
\tau_{\text{D}}$) = $S_0\exp{
(-\tau_{\text{D}}/T_1)}+S_1$.  
  Note that these  
$T$\,$<$\,$T_Z$ relaxation rates are faster
than the rate at $T$\,$\approx$\,$T_Z$ for $\nu$=1 
\cite{science}.
  Qualitatively, this
shows that there is a greater overlap of the 
density of states for electrons with opposite 
spins at $\nu$=$\frac{1}{2}$ than at $\nu$=1.

The isotropic Fermi contact hyperfine coupling 
between the electron spins and the nuclear spins 
is responsible for both $K_S$ and 
1/T$_1$ \cite{pankaj,nick,opnmr,science}, as is the case for some 
metals\cite{slichter,winter}. Within the WICFM,
1/T$_1$($\nu$=$\frac{1}{2}$,$T$) for $^{71}$Ga nuclei 
in the center of the quantum well
is :
\begin{equation}  
\frac{1}{\text{T}_1}=\frac{\pi (m^{\ast})^2}{\hbar^3}
\Bigl(\frac{K_{S\text{int}}^{{\cal P}=1}}{n}\Bigr)^2 
\frac{ k_{\text{B}} 
T}{1+\exp{[\frac{\mu(T)}{k_BT}(\delta(T)-1)]}}.
\label{eq5}
\end{equation}  

This expression is used as a two-parameter fit
to the 1/T$_1$($T$) data (Fig.\,\ref{fig3},
solid lines), where $\mu(T)$ and $\delta(T)$ are
obtained from  Eqns.\,2--4  for each $m^{\ast}$ and $J$.
The behavior of 1/T$_1$$(T)$ as 
$T\!\rightarrow0$ is also
quite sensitive to the parameter $\delta(0)$.
In Fig.\,\ref{fig3}, we illustrate $\delta(0) < 1$ 
with the dotted curve, $\delta(0)$\,=\,1 with the 
dashed curve, and we find $\delta(0)$\,$>$\,1 
for Cases I and II (solid curves).  In 
contrast to normal metals, here
$\mu(0)$$\sim$$|$g$^{\ast}\mu_eB_{
\text{tot}}|$.

\begin{figure}
\centerline{\epsfxsize=3.23in\epsfbox{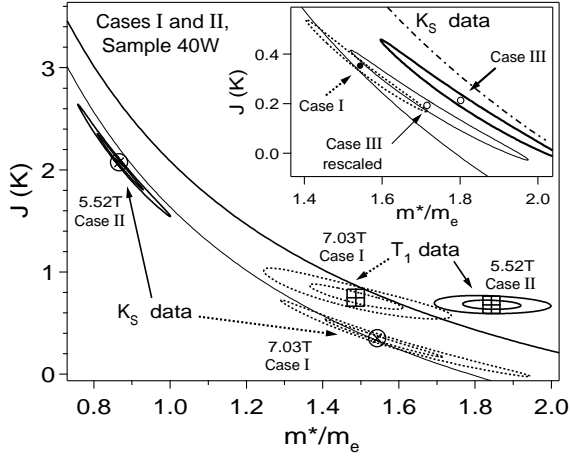}}
\caption{ Values of $J$ (in Kelvin) and $m^{\ast}$ (in
units of the  electron mass in vacuum, $m_{e}$),
 obtained using  
a $\chi$$^2$ analysis  of :  (main)\, 
 ${\cal P}(T)$ (circles) and 1/T$_1(T)$ (squares)
for Case I
(dashed contours) and Case II 
(solid contours), and  (inset) ${\cal P}(T)$  
for Case I (dashed contour) and Case III (thick solid
 contour).  The thin solid contour in the inset
shows the rescaled Case III values described in the text.
Within the WICFM, the ground state is fully spin-polarized
for ($J$,$m^{\ast}$) values that lie above the thin solid curve (Case I),
thick solid curve (Case II), or dashed-dotted curve (Case III,
inset).}
\label{fig4}
\end{figure}

 Figure \ref{fig4} shows the best values of
$J$ and $m^{\ast}$ obtained for each data set
in Figs.\,\ref{fig2}(b,c) and \ref{fig3}.
The correlation between $J$ and $m^{\ast}$
is shown by $\Delta$$\chi$$^2$=1 and
$\Delta$$\chi$$^2$=4 contours.
These  ($J$,$m^{\ast}$) values   lie
quite close to the curves which mark the 
transition between
fully and partially polarized ground states 
(i.e., where $\delta(0)$=1). 
There is negligible overlap between the 
contours and the line $J$=0,
so the non-interacting composite
fermion model used by Kukushkin et al.
\cite{kukushkin} does not work here. 
 Moreover, there is no ($J$,$m^{\ast}$)
pair which can simultaneously describe
the four data sets measured using the
same sample and $B_{\perp}$ (Fig. 4 (main)),
 so we conclude that 
even the weakly-interacting composite
fermion model is a poor description of the 
$\nu$=$\frac{1}{2}$ state for these Cases.  
The most glaring inconsistency is that of
Case II, where $\delta(0)$$<$1 (i.e., 
partially-spin polarized at $T$=$0$)
is inferred from ${\cal P}(T)$, which is 
{\em incompatible} with the
result $\delta(0)$$>$1 (i.e., fully-spin 
polarized at $T$=0) that is inferred
from 1/T$_1$$(T)$.

Figure \ref{fig4} (inset) shows the
($J$,$m^{\ast}$)
values obtained from the 
 ${\cal P}(T)$ data for Cases I and III.  
These
values do not agree,  however, sample 10W and 
40W also have slightly different
electron densities and well widths ($n$,$w$). 
 This 
 would affect our 
results, since we expect  $k_BJ$ $\propto$ 
$eB_{\perp}/m^{\ast}$ $\propto$ 
$E_C(\lambda)\equiv e^2/(\epsilon\sqrt{l_0^2+\lambda^2})$, where 
$l_0$=$\sqrt{\hbar c/eB_{\perp}}$ is the 
magnetic length, $\epsilon$=13, and the
parameter $\lambda\approx \frac{1}{4} w$ 
modifies the Coulomb energy
scale due to the non-zero thickness $w$ of 
the quantum well \cite{zhang}.
To correct for this, the Case III ($J$,$m^{\ast}$) 
values are rescaled using:
\begin{eqnarray}
\frac{J(n_{\text{I}},w_{\text{I}})}{J(n_{\text{III}},w_{\text{III}})} = 
\sqrt{\frac{n_{\text{I}}}{n_{\text{III}}}
\Bigl(\frac{1+\frac{n_{\text{I}}w_{\text{I}}^2}
{n_{\text{III}}w_{\text{III}}^2}}{2}\Bigr)}\,,\nonumber\\
\frac{m^{\ast}(n_{\text{I}},w_{\text{I}})}{m^{\ast}(n_{\text{III}},w_{\text{III}})} =
 \sqrt{\frac{n_{\text{I}}}{n_{\text{III}}}\Bigl(\frac{2}{1+
\frac{n_{\text{I}}w_{\text{I}}^2}{n_{\text{III}}w_{\text{III}}^2}}\Bigr)}\,.
\end{eqnarray}
The rescaled contour has a good overlap 
with the ($J$,$m^{\ast}$) values
for Case  I (Fig.\,\ref{fig4} (inset)).
This rescaling is irrelevant for Fig.\,\ref{fig4} (main),
 where the results on a single sample are shown.

In conclusion, neither a non-interacting 
nor a weakly-interacting
composite fermion model is sufficient to 
explain our experiments,
which probe the electron spin degree of 
freedom {\em right at} $\nu$=$\frac {1}{2}$.  
Knight shift and 1/T$_1$ data, taken together,
 provide important new 
constraints on the theoretical description
of the $\nu=\frac {1}{2}$ state.  Finally, 
in addition to fully polarized ground
states (Case~I, $\frac{k_BT_Z}{E_C(\lambda)}$=0.021),
partially spin-polarized
ground states (Cases II and III, 
$\frac{k_BT_Z}{E_C(\lambda)}$=0.017 
and 0.019) are 
experimentally accessible, and 
exhibit unexpected features (e.g., 
the extra linewidth).

We thank  
N.\ Read, R.\ Shankar, S.\ Sachdev, and 
R.\,L.~Willett for many helpful
discussions.   This work was
supported by NSF Grant $\#$DMR-9807184.  
SEB also acknowledges an Alfred P.\ Sloan
Research Fellowship.


\begin{references}
\bibitem{cfrevs}For comprehensive reviews, see:  
     {\em Composite Fermions}, ed. by
     O.\,Heinonen (World Scientific, 1998);
     {\em Perspectives in Quantum 
     Hall Effects}, ed.\, by S.\,Das Sarma and 
     A.~Pinczuk (Wiley, New York, 1997);
     J.\,K.\ Jain, Adv.\ Phys.\ {\bf 41}, 105 (1992);
     R.\,L.\ Willett, {\em ibid.} {\bf 46}, 447 (1997).
\bibitem{rectheory}For recent theoretical developments near
     $\nu$=$\frac{1}{2}$, see: N.~Read, 
     Phys.\ Rev.\ B {\bf 58}, 16262 (1998);
     A.\ Stern, B.\,I.\ Halperin, F.\,v.\,Oppen,
     and S.\,H.\ Simon, {\em ibid.} {\bf 59}, 12547 (1999);
     R. Shankar, cond-mat/9903064.
\bibitem{HLR}B.\,I.\ Halperin, P.\,A.\ Lee, and 
     N.\ Read, Phys.\ Rev.\ B {\bf 47}, 7312 (1993).
\bibitem{willett93}R.\,L.\ Willett, R.\ R.\ Ruel, K.\ W.\ West,
     and L.\ N.\ Pfeiffer, 
     Phys.\ Rev.\ Lett.\ {\bf 71}, 3846 (1993).
\bibitem{willett95}R.\,L.\ Willett, K.\,W.\ West,
     and L.\,N. Pfeiffer, Phys.\ Rev.\ Lett.\
     {\bf 75}, 2988 (1995).
\bibitem{kang}W.\ Kang {\em et al.}, 
     Phys.\ Rev.\ Lett.\ {\bf 71}, 3850 (1993).
\bibitem{goldman94}V.\,J.\ Goldman, B.\ Su, and J.\,K.\ Jain, 
     Phys.\ Rev.\ Lett.\ {\bf 72}, 2065 (1994).
\bibitem{smet}J.\,H.\ Smet {\em et al.},
     Phys.\ Rev.\ Lett.\ {\bf 77}, 2272 (1996).
\bibitem{du}R.\,R.\ Du {\em et al.},
     Phys.\ Rev.\ Lett.\ {\bf 75}, 3926 (1995).
\bibitem{kukushkin}I.\,V.\ Kukushkin, K.\,v.\,Klitzing,
     and K.\ Eberl,
     Phys.\ Rev.\ Lett.\ {\bf 82}, 3665 (1999).
\bibitem{polarmass}K.\ Park and J.\,K.\ Jain,
     Phys.\ Rev.\ Lett.\ {\bf 80}, 4237 (1998).
\bibitem{opnmrchar}S.\,E. Barrett, R. Tycko, 
     L.\,N.\ Pfeiffer, and K.\,W.\ West, 
     Phys.\ Rev.\ Lett.\ {\bf 72}, 1368 (1994).
\bibitem{pankaj}P.\ Khandelwal {\em et al.},
     Phys.\ Rev.\ Lett.\ {\bf 81}, 673 (1998);
     also available at cond-mat/9801199.
\bibitem{nick}N.\,N.\ Kuzma {\em et al.},
     Science {\bf 281}, 686 (1998); 
     also available at cond-mat/9907279.
\bibitem{insensitive}L.\,N.\ Pfeiffer {\em et al.}, 
     Appl.\,Phys.\,Lett. {\bf 61},\,1211\,(1992). 
\bibitem{opnmr}S.\,E.\ Barrett {\em et al.},
     Phys.\ Rev.\ Lett.\ {\bf 74}, 5112 (1995).
\bibitem{science}R. Tycko {\em et al.},
     Science {\bf 268}, 1460 (1995).
\bibitem{slichter}C.\,P.\ Slichter, {\em Principles 
     of Magnetic Resonance} (Springer, New York, 
     1990), 3$^{\text{rd}}$ ed.
\bibitem{density}In simulations, 
     we found that an unreasonably large FWHM of 
     70\% or more
     is required for a gaussian distribution of electron
     densities along the well to explain the large
     temperature independent linewidth observed 
     at $\nu$=$\frac{1}{2}$ for Cases II and III.
\bibitem{winter}J.\ Winter,  
     {\em Magnetic Resonance in Metals},
     (Oxford Univ.\,Press, London, 1971). 
\bibitem{zhang} F.\,C.\ Zhang and S.\ Das Sarma,
     Phys.\ Rev.\ B {\bf 33}, 2903 (1986).


\end{references}
  \end{document}